\documentclass[epj,twocolumn]{webofc}
\usepackage[varg]{txfonts}   % Web of Conferences font
\woctitle{Hadron Collider Physics symposium 2012}
\begin{document}
\title{Single Top Physics at Hadron Colliders}

\author{Giorgio 
Chiarelli\inst{1}\fnsep\thanks{\email{giorgio.chiarelli@pi.infn.it }} 
        % etc.
\\
on behalf of the ATLAS, CDF, CMS, D0 Collaborations
}

\institute{Istituto Nazionale di Fisica Nucleare, Sezione di Pisa Via 
B.~Pontecorvo 3, I-56127 Pisa
          }

\abstract{%
The production of top quark in electroweak processes was first observed in 
2010 at the Tevatron. Since then it has been carefully studied at both LHC 
and Tevatron. Single top production proceeds through different channels 
and allows a direct determination of Cabibbo-Kobayashi-Maskawa matrix 
element $|V_{tb}|$. We will present the current status of searches and 
observation and discuss the results obtained so far and perspectives at 
hadron machines.
} 
\maketitle
\section{Introduction}
\label{intro}
\begin{figure}
\centering
\includegraphics[width=8cm,clip]{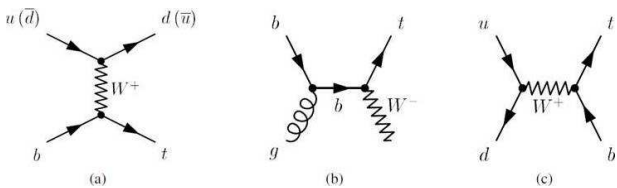}
\caption{Feynman diagrams for single top production}
\label{feynm}       % Give a unique label
\end{figure}

Top quark is produced in pair in strong interactions but can also be 
produced in electroweak interactions. In the latter processes only one top 
quark appears in the final state. Therefore this set of processes is 
generically dubbed {\it single top}. Formally the single top production 
proceeds by different Feynman diagrams and 
you can distinguish in figure~\ref{feynm} production through $t$~(a), 
$Wt$~(b) 
and $s$~(c) channels.
At hadron colliders (see table~\ref{xse}) the three processes have quite 
different 
cross sections strongly dependent by the center of mass energy and the PDF 
of the incoming partons.
Going from 2 to 7 and 8 TeV, $t$ and $Wt$ production increases by factor 
$30\div40$ while $s$ channel only increases by a factor 4. We will discuss 
later the implications of this effect, due to the different ($pp$ versus 
$p\bar{p}$) initial state.

Historically, while top quark was discovered in 1995~\cite{disc} 
in $p\bar{p} \rightarrow t\bar{t}$ process, 
due to the very low signal(S)-over-background(B) ratio
it took 
14 more years to see single top at the Tevatron~\cite{cdfd},~\cite{d0d}.
ATLAS and CMS, instead, thanks to the much larger cross 
sections and better S/B available at the LHC, 
already saw single top in the $t$ channel in 2011 
and measured its cross section in 2012~\cite{slhcatl},~\cite{slhcms}. 

$Wt$ channel, completely impossible to identify at the Tevatron, was 
explored 
and recently evidence was reported by both ATLAS~\cite{atlwt} and 
CMS~\cite{cmswt}.  
\subsection{Physics interest}
\label{intro-2}
The main interest in measuring single top properties  is, set aside 
comparison to prediction, that its production cross section is 
proportional to
the CKM matrix element $|V_{tb}|$ under the assumption that 
$|V_{tb}|>>|V_{ts}|, |V_{td}|$. Actually you measure $|f_L\times 
V_{tb}|$ 
where $f_L=1$ within the Standard Model. In other words, directly 
measuring $|V_{tb}|$ we can test models with more than three 
generations, anomalous couplings in top sector and other new physics 
scenarios.

\section{Signal and Background}
\label{sign}
The single top final state is characterized by at least a $W$ and 2 or 3 
jets. One jet 
and the $W$ come from the top decay, while the other jets are related to 
the production channel. Therefore in $t$ channel the final state has  
2 or 3 jets 
(one, not coming from top, mostly in the forward region), in the $s$ 
channel two jets 
from $b$ fragmentation. Finally the $Wt$ channel has two $W$s and one jet 
in the final state.
The topology for $s$ and $t$ channel strongly resembles the $WH$ 
associate production 
making the observation of this process important for SM Higgs 
studies as well as for searches of new bosons ($W^{'}$, $H^+$).
Due to the need to suppress background, events are collected using 
high-$P_T$ lepton triggers, therefore the final states that will be 
used always contain $W$s decaying into leptons.
\begin{table}
\centering
\caption{Single top production cross sections (in pb)~\cite{theo}.}
\label{xse}       % Give a unique label
% For LaTeX tables you can use
\begin{tabular}{llll}
\hline
Energy&$t$ channel&$s$ channel&$Wt$ channel\\\hline
\hline
Tevatron (1.96 TeV) & $2.26\pm 0.2$ & $1.04\pm0.1$&$0.3\pm0.06$ \\
LHC (7 TeV) & $64.2\pm2.4$ & $4.6\pm 0.2$&$15.7\pm 1.1$ \\
LHC (8 TeV) & $87.8\pm3.4$ & $5.6\pm 0.3$&$22.4 \pm 1.5$\\\hline
\end{tabular}
% Or use
%\vspace*{5cm}  % with the correct table height
\end{table}

From the topologies described it is obvious that any experiment looking 
for single top must identify charged ($e$, $\mu$, 
$\tau$) and neutral leptons, latter through measurement of missing 
transverse energy (from now on MET). It must also be able to reconstruct 
jets and tag the ones containing heavy flavour. Since the last decade of 
the past century any experiment at hadron colliders has such capabilities.
\subsection{Backgrounds}
\label{back}
In a final state containing a $W$ and two or three jets, the background is 
composed of:
\begin{itemize}
\item $t\bar{t}$ events from strong interactions, $WW$, $WZ$, $ZZ$ 
events, $Z+$ heavy flavours ($HF$), $Z+$ light flavour ($lf$) estimated 
using MC 
simulations;
\item multijet ("QCD") events where one jet fakes a lepton and a 
mismeasurement of transverse energy generates a fake neutrino signal;
\item $W+HF$ ($W$+heavy flavour jets), $W+lf$ ($W$+light flavour jets).
\end{itemize}

Multijet contribution, being due to a 
combination of very low probability topology of events with large yield 
together with non-well modeled
detector features, is measured in data. This procedure, however,
carries a large uncertainty (usually $30\div 40 \%$).
$W+lf$ and $W+HF$ contributions are computed by algorithms that 
use a mixture of data-driven and Monte Carlo estimates.

In absence of a final state containing any heavy flavour, a $b$ tag can 
be faked by failure of tracking and/or pattern recognition algorithms. 
Fakes are parametrized as a function of jet and event characteristics 
using 
control 
data samples.
\section{Selection}
\label{sele}
After requiring a high $P_T$ charged lepton, large MET, and two high 
$E_T$ jets, all experiments apply specific requirements to suppress 
multijet background. Useful quantities are the angular distance ($\Delta 
\phi$) between physics objects, the scalar sum of the $E_T$ of the visible 
objects ($H_T$), the transverse mass ($M_{T}^{W}$) between the charged and 
the neutral leptons and of course the MET.

CDF, ATLAS and CMS directly use a cut in the 
MET-$M^{W}_{T}$ plane, D0 applies a cut in $H_T$ and in $\Delta \phi(MET, 
l)$. No attempt is made to optimize rejection of this background using 
multivariate techniques employed in other analyses.

Furthermore ATLAS and CMS exploit the fraction of jet tracks pointing 
to the primary vertex to reject background events due to pileup.

\section{Signal Extraction}
\label{signale}
After all requirements the situation in the two accelerators is 
quite different. At the Tevatron the signal is still swamped by background 
(see table~\ref{d0bck}) while at the LHC, signal is  
$\approx 50\%$ of the $t\bar{t}$ background ($t$ channel). $s$ channel is 
only 5\% of the $t$-channel signal while $Wt$ is $\approx 50\%$ of the 
$t$-channel (see for example table~\ref{atlbck}). Therefore the strategies 
followed by the Tevatron experiments 
are different from the ones at the LHC. 
 \subsection{CDF and D0}
\label{tev}
Due to the tiny yields and low S/B historically Tevatron experiments 
did not attempt to distinguish between $s$ and $t$ 
contributions but rather to establish the existence of the signal.
\begin{table}
\centering
\caption{Sample composition in 5.4 fb$^{1}$ of data, after selection, at 
D0.}
\label{d0bck}       % Give a unique label
% For LaTeX tables you can use
\begin{tabular}{llll}
\hline
Source&2-jets&3-jets&4-jets\\\hline
\hline
$tb$& $104\pm 16$ & $44\pm7.8$&$13\pm3.5$ \\
$tqb$& $140\pm13$ & $72\pm 9.4$&$26\pm 6.4$ \\
$t\bar{t}$& $433\pm87$ & $830\pm 133$&$860 \pm 163$\\
$W+jets$& $3560\pm354$ & $1099\pm 169$&$284 \pm 76$\\
$Z+jets$,multiboson& $400\pm55$ & $142\pm 41$&$35\pm 18$\\
Multijet& $277\pm34$ & $130\pm 17$&$43 \pm 5.2$\\\hline
Total expectation&$4914\pm558$&$2317\pm377$&$1261\pm272$\\
\hline
Data&4881&2307&1283\\\hline
\end{tabular}
% Or use
%\vspace*{5cm}  % with the correct table height
\end{table}
For that reason the analyses we present here follow the same strategy: 
first 
measure the $s+t$ cross section combined and then attempt to determine $s$ 
and $t$ channels separately.
The low S/B ($\approx 1/20$) makes mandatory the use of multivariate 
analysis tools to identify the single top contribution. 

\begin{figure}
\centering
\includegraphics[width=8cm,clip]{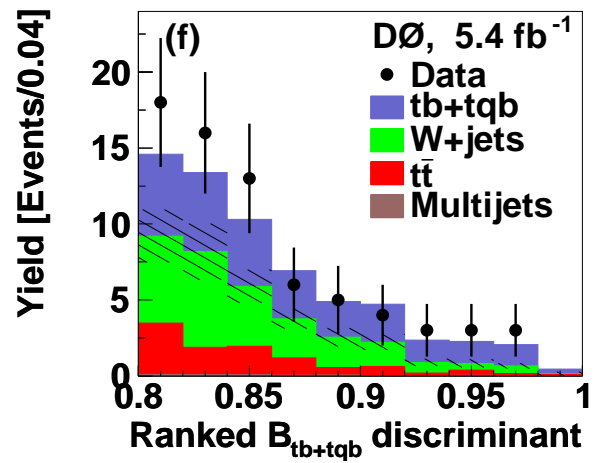}
\caption{Final BNN discriminant for single top search at D0.}
\label{d0output}       % Give a unique label
\end{figure}

D0 uses three different multivariate techniques (Bayesian Neural Network, 
Boosted Decision Tree, NEAT). Each algorithm is trained and 
optimized for $s$ and $t$ channel separately. Their output is fed to a 
final Bayesian Neural Network (BNN) discriminant. The output of the BNN is 
shown in figure~\ref{d0output} for the signal rich (>0.8) region.
By fitting the various components of the BNN output, D0 measures, 
with an integrated luminosity of 5.4 $fb^{-1}$, a 
cross section $s+t$ of $3.43 \pm 0.74$ pb.

As a next step one can fix the $s/t$ ratio to its Standard Model value and 
find a value for the cross section of $\sigma(t)=2.96\pm 0.66$ pb and 
$\sigma(s)=0.68\pm 0.36$ pb.
In turn this result can be used to set a limit $|f_L\times V_{tb}|>0.79$ 
at 95 \%C.L.~\cite{d0res}.

By leaving the two cross section free D0 measures $\sigma(t)=2.9\pm 0.59$ 
pb and $\sigma(s)=0.98\pm 0.63$~\cite{d0res1}.
Obviously the latter result does not 
represent an evidence for $s$ production.

After presenting the first observation using a data set corresponding to 
3.4 $fb^{-1}$ and three different analysis techniques, CDF updated one of 
these analyses, based on the use of a set of Artificial Neural 
Networks (ANN) discriminants separately 
optimized for the $s$ and $t$ channel and for samples with 2 and 3 jets 
using a data set of 7.5 $fb^{-1}$. With respect to the original analysis 
several improvements were applied: increased acceptance for $e$ and $\mu$, 
use of additional trigger paths, a new QCD 
background suppression technique. 
Finally an improved simulation of the signal (using POWHEG 
generator)~\cite{manfredi}).
After selection, the outputs of the different discriminants are used as 
input to a final ANN optimized for $s+t$~\cite{cdfd}. 

\begin{figure}
\centering
\includegraphics[width=8cm,clip]{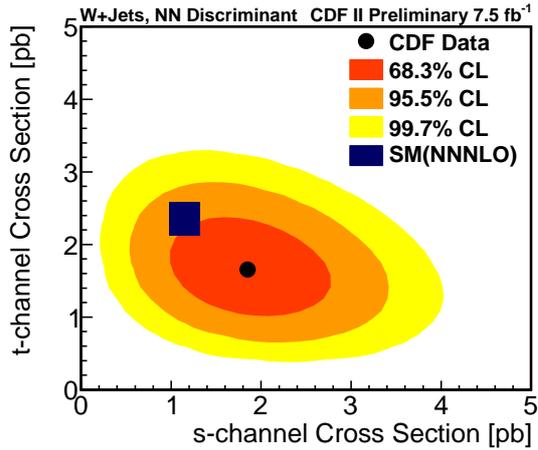}
\caption{CDF result for 7.4 fb$^{-1}$ compared to SM prediction. 
$\sigma(t)$ and $\sigma(s)$ are 
fit independently.}
\label{2dcdf}       % Give a unique label
\end{figure}

The result is $\sigma(s+t)=3.04^{+0.57}_{-0.53}$, 
and $|V_{tb}|>0.75$ at 95 \% C.L.
Finally CDF fits the two channels simultaneously and obtains (see 
figure~\ref{2dcdf}) 
$\sigma(s)=1.81^{+0.63}_{-0.58}$ pb and $\sigma(t)=1.49^{+0.63}_{-0.58}$ 
pb~\cite{cdf75}. This result is in agreement with SM expectations at 
the
$\approx 1 \ \sigma$ level.  

\subsection{ATLAS and CMS}
\label{lhc}
The large amount of data available at the LHC, combined with the increase 
in cross section, allow a thorough study of many top properties 
and production mechanisms. However, as already mentioned, the increase in 
background makes the $s$ channel very difficult to observe. Therefore both 
Collaborations, so far, concentrated on the $t$ and $Wt$ channel.

For $t$ channel the strategy (common to both experiments) is to split the 
lepton+jets sample
by jet number (2,3) and number of $b$-tags (0,1,2). The signal region 
is the one with two jets and one tag, the other two are used as control 
regions.
Table~\ref{atlbck} shows, as an example, the sample composition after all 
requirements in the  8 TeV ATLAS data.
In the 2 jet bin
$t$ channel is $\approx$ 12 \% of the total number of events, $Wt \approx 
$ 4 \% and $s$ channel is less than 1 \%.
\begin{table}
\centering
\caption{Sample composition after selection in ATLAS data set (5.8 
fb$^{-1}$) at 8 TeV.}
\label{atlbck}       % Give a unique label
% For LaTeX tables you can use
\begin{tabular}{lll}
\hline
Source&2-jets&3-jets\\\hline
\hline
$t$-channel& $5210\pm 210$ & $1959\pm78$\\
$s$-channel& $343\pm14$ & $100\pm 4$\\
$Wt$& $1570\pm110$ & $1363\pm 95$\\
$t\bar{t}$& $11700\pm1200$ & $15300\pm 1500$\\
$W+light flavor$& $5500\pm1700$ & $1160\pm 350$\\
$W+heavy flavor$& $12000\pm6000$ & $3900\pm 2000$\\
$Z+jets$,multiboson& $1200\pm720$ & $410\pm 240$\\
Multijet& $3000\pm1500$ & $1650\pm 830$\\
\hline
Total expectation&$41600\pm 6600$&$25800\pm2700$\\
\hline
Data&40663&23697\\\hline
\end{tabular}
% Or use
%\vspace*{5cm}  % with the correct table height
\end{table}

\begin{figure}
\centering
\includegraphics[width=8cm,clip]{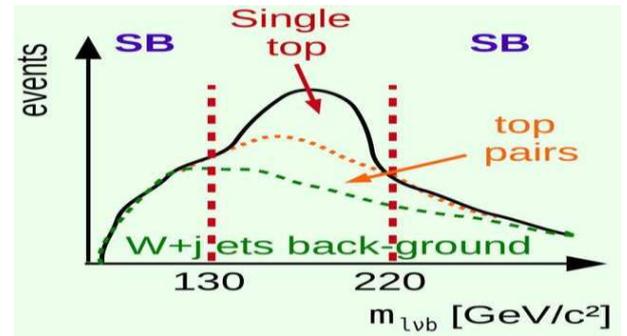}
\caption{Schematic view of the $M_{l\nu b}$ variable showing signal and 
control region.}
\label{mlnb}       % Give a unique label
\end{figure}
CMS explored the data set collected in 2011 at 7 TeV
searching for top produced 
in $t$ channel using three different techniques.
One (dubbed "$\eta_{j'}$") exploits a characteristic feature of 
$t$-channel single top events: the forward pseudorapidity of the light jet 
recoiling against the top. 
The other two (Neural Network and BDT) are multivariate techniques 
that compare signal expectation with observation. 
The aim of the multivariate analyses is to combine the information 
contained in the
signal enriched and signal depleted regions to obtain a precision 
measurement of the single top cross section. The luminosity used is 
$\simeq 1.2\,(1.7)$ fb$^{-1}$ for electrons(muons).

\begin{figure}
\centering
\includegraphics[width=8cm,clip]{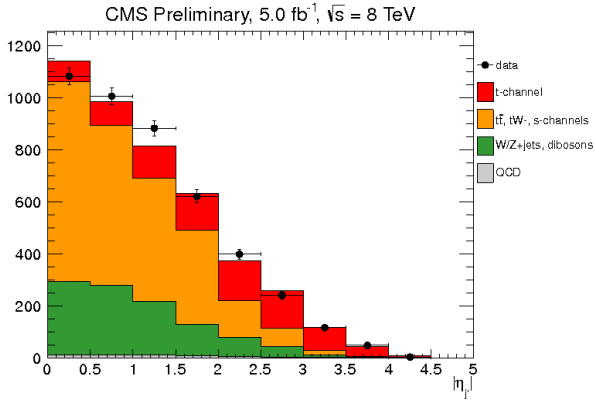}
\caption{CMS data: $|\eta_{j'}|$ distribution in the single-top 
enriched 
region.}
\label{etap}       % Give a unique label
\end{figure}

In the $\eta_{j'}$ analysis 
CMS exploits the different kinematical properties of signal and $W+$ jets 
background to further split the signal region (2 jet, 1 tag) into signal 
rich and background enriched regions.
In figure~\ref{mlnb} you can see the mass distribution for the $M_{l\nu 
b}$ system~\cite{jean}. 
After requiring $130<M_{l\nu b}<220$ GeV/c$^2$ the S/B 
reaches $\simeq 1/5$. This provides
enough separation that a simple fit to the 
$|\eta|$ distribution of the most forward jets is sufficient to measure 
the single top cross section (see figure~\ref{etap}).

The results of the three analyses are the following:
$\eta_{j'}$: $\sigma(t)=70\pm 6(stat)\pm 6.5 (syst) \pm 3.6 (theo) \pm 
1.5 (lumi)$~pb; NN: 
$\sigma(t)=68.1\pm 4.1(stat)\pm 3.4 (syst)^{+3.3}_{4.3}(theo) \pm 1.5 
(lumi)$~pb; BDT:
$\sigma(t)=66.6\pm 4(stat)\pm 3.3 (syst)^{+3.9}_{-3.3}(theo) \pm 1.5 
(lumi)$~pb.

All results are consistent with SM expectations and among each other.

By combining the three measurements,
the overall result is
$\sigma(t)=67\pm 4(stat)\pm 3 (syst) \pm 4 (theo) \pm 2 (lumi)$~pb
and $V_{tb}=1.02\pm 0.05 \pm 0.02(theo)$ or $|V_{tb}|>0.92$ at 95 \% 
C.L.~\cite{slhcms}. 

In the 8 TeV sample more stringent 
requirements are applied in order to deal with the pile-up. 
While at 7 TeV the typical number was $5\div 10$ events per crossing,
the increase in luminosity brought this number in 2012 up to well above 20
events/crossing (with a recording by CMS of 78 events in a given 
crossing)~\cite{neu}. 
Despite this additional challenge
CMS measures, using just the $|\eta_{j'}|$ 
technique, 
$\sigma(t)=80\pm 6(stat)\pm 11 (syst) \pm 4 (lumi)$~pb
and $V_{tb}=0.96\pm 0.08 \pm 0.02(theo)$ or $V_{tb}>0.81$ at 95 \% 
C.L.~\cite{cms8}.

\begin{figure}
\centering
\includegraphics[width=8cm,clip]{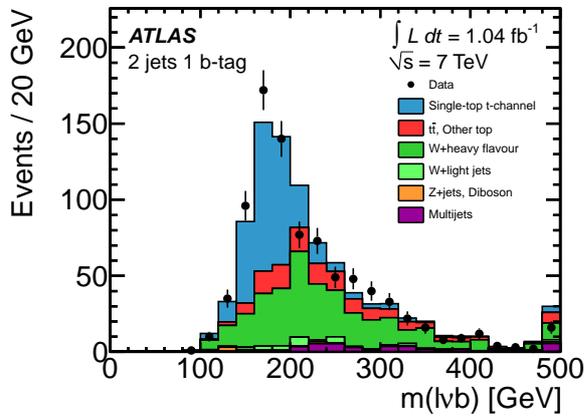}
\caption{$M_{l \nu b}$ distribution in ATLAS 7 TeV data set.}
\label{atlmln}       % Give a unique label
\end{figure}

Production through $t$ channel was successfully observed at 7 TeV also by 
ATLAS. 
Key is, again, the selection of events where a $W$ is produced in 
association with 2 or 3 jets and one $b$-tag (signal region) or no tag 
(control region). After selection S/B $\sim 
1/9$ and an Artificial Neural Net is used. In figure~\ref{atlmln} 
the $M_{l\nu b}$~distribution (one of the input to the ANN) is shown for 
the 2 jet events sample. Signal is clearly visible. 
Fitting the output 
discriminant, ATLAS measures 
$\sigma(t)=83\pm 4(stat) ^{+20}_{-19} (syst)$~pb
and $V_{tb}=1.13\pm 0.14 (stat+sys)$ or $V_{tb}>0.75$ at 95 \% 
C.L.~\cite{slhcatl}.
Splitting the sample in $t$ and $\bar{t}$, ATLAS measures
$\sigma(t)=53\pm 10.8 (syst)$~pb and
$\sigma(\bar{t})=29.5 \pm 7.5 (syst)$~pb in agreement with SM prediction.
The ratio $R$ of these two cross sections can be used to constrain the 
ratio of $u/d$ PDFs. ATLAS measures $R=1.81^{+0.33}_{-0.22}$ in agreement 
with predictions~\cite{atlas7}.

At 8 TeV the signal growth 
by 35 \% is 
matched by $t\bar{t}$ increase (40 \%) and $W+jets$ ($25\div 35$ \%). What 
is most challenging is, as already mentioned, the pileup.
The growing number of overlapping events requires 
harder lepton isolation cuts, and re-tune dedicated 
requirements in order to keep 
it under control. In the end S/B$\sim 1/8$ (see 
table~\ref{atlbck}) and, applying the same 
analysis technique used for the 7 TeV data set, ATLAS measures
$\sigma(t)=95\pm 2(stat)\pm 18 (syst)$~pb
and $V_{tb}>0.80 $ at 95 \% C.L.~\cite{atlas8}.

\begin{figure}
\centering
\includegraphics[width=6cm,clip]{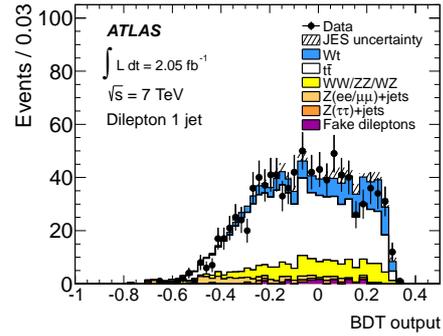}
\caption{ATLAS search for single top in $Wt$ channel: 
BDT output in the $N_{jet}=1$ sample. $Wt$ contribution normalized to SM 
prediction.
}
\label{atlbdtwt}       % Give a unique label
\end{figure}

\section{Single top in $Wt$ channel}
\label{wt}
While $Wt$ production has a negligible cross section at the Tevatron,its 
yield at the LHC allows to look for it. Both experiments 
searched for $Wt$ in the 7 TeV data using the dilepton channel (when both 
the primary $W$ and the one from $t$ decay leptonically).

In table~\ref{bckwt} we show the ATLAS sample composition in the 
$Wt\rightarrow l\nu l\nu b$ 
sample (with $l=e, \mu$). 
Despite the large $t\bar{t}$ background, the 
$N_{jet}=1$ bin shows a S/B $\approx$ 18 \%.
Therefore ATLAS, using a BDT with 22 input variables, is able to
see an evidence 
with a significance of $3.3 \sigma$ (3.4 expected) in a data sample
at 7 TeV corresponding to an
integrated luminosity of 2.05 fb$^{-1}$ (see figure~\ref{atlbdtwt}).
The 
cross section is
$\sigma(Wt)=16.8\pm 2.9(stat) \pm 2.9 (syst) $~pb from which  
$|V_{tb}|=1.03^{+0.16}_{-0.19}$~\cite{atlwt}.
\begin{table}
\centering
\caption{Sample composition after selection in ATLAS data set (2.05
fb$^{-1}$) at 7 TeV.}
\label{bckwt}       % Give a unique label
% For LaTeX tables you can use
\begin{tabular}{llll}
\hline
Source&1-jet &2-jet&$\ge$ 3-jet\\\hline
\hline
$Wt$& $147\pm13$ & $60\pm 9$&$17\pm5$\\
$t\bar{t}$& $610\pm110$ & $1160\pm 140$&$740\pm30$\\

Diboson& $130\pm17$ & $47\pm5$&$17\pm 4$\\
$Z\rightarrow ee$& $20\pm 2$ & $11\pm 2$&$5\pm 2$\\
$Z\rightarrow \mu \mu$& $29\pm 3$ & $28\pm 3$&$12\pm 3$\\
$Z\rightarrow \tau \tau$& $9\pm 6$ & $4\pm 4$&$2\pm 1$\\
Fake dileptons& $11\pm11$ & $5\pm5$&negl.\\
Total backg.& $810\pm120$ & $1260\pm140$&$780\pm 130$\\
\hline
Total expectation&$960\pm 120$&$1320\pm140$&$790\pm 130$\\
\hline
Data Observed&934&1300&825\\\hline
\end{tabular}
\end{table}

\begin{figure}
\centering
\includegraphics[width=6cm,clip]{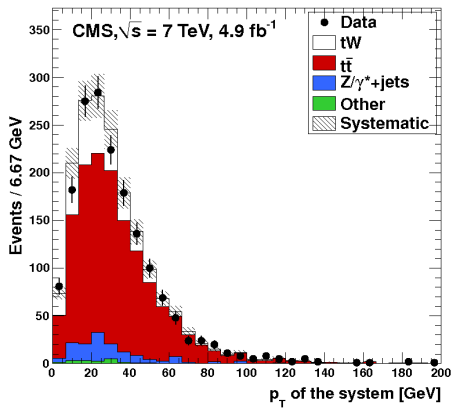}
\caption{CMS search for single-top 
in $Wt$ channel: $P_T$ of the system composed by the two 
leptons, MET and the leading jet. 
}
\label{ptwt}       % Give a unique label
\end{figure}

CMS follows the same strategy. The signal region 
is defined by one jet, 1 b-tag while control regions are defined as 
2 jets, 1 or 2 b-tags. S/B is 24 \% in the signal region.
Input to a BDT are four variables: $H_T$ (scalar sum of the transverse 
energy of the visible objects), the $P_T$ of the leading jet, the angular 
distance between the MET and the closest lepton and
the $P_T$ of 
the system composed by the two leptons, MET and the 
jet (signal region) (shown in fig.~\ref{ptwt}).
 
CMS obtains an evidence of 4 
$\sigma$ and measures, in a data set of 4.9 fb$^{-1}$, $\sigma(Wt)=16.\pm 
5(stat) \pm 4 (syst) $~pb and extracts 
$|V_{tb}|=1.01^{+0.16}_{-0.13}(stat+syst)\pm 0.03(theo)$~\cite{cmswt}.
 
\section{Conclusion}
By Fall 2012 the production of single top is definitively well 
established at least in two of the three channels.

\begin{figure}
\centering
\includegraphics[width=9cm,clip]{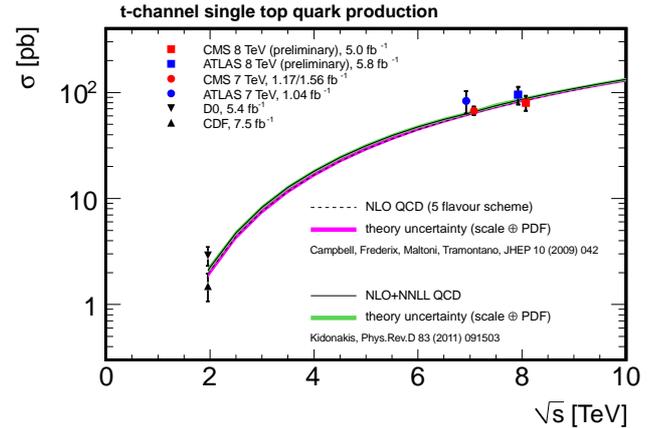}
\caption{Single top cross section at Tevatron and LHC($t$ 
channel only), theoretical prediction and actual measurements.}
\label{txsec}       % Give a unique label
\end{figure}

Figure~\ref{txsec} summarizes the $t$ channel measurements at different 
energies 
and different colliders. The agreement with theoretical prediction is
striking. 

Production through $Wt$, thanks to the large statistics accumulated at the 
LHC, is established at the 4$\sigma$ level. Given that the measurement is 
by far statistics limited, the expectation is that will 
reach the 
5 $\sigma$ level for a full "observation" by Winter 2013 once the 8 TeV 
data set are analyzed.

Production through $s$ channel, however, is not yet fully established. CDF 
measures $\sigma(s)=1.81^{+0.63}_{-0.58}$ with two-third of its data set.
While the Tevatron is by now closed, 
analyses are in progress 
to update single top cross section measurements with the whole data set.
Besides, $s$ channel specific strategies and the use of new $b$ tagging 
tools are being pursued.

At the LHC, despite the large statistics, currently the very low S/B makes
$s$ channel a process difficult to observe. Dedicated strategies 
must still be developed.
In general the increase of luminosity comes with a dear price of 
a very large amount of pile-up. While the Collaborations have shown their 
capability to deal with this new challenge, its impact on the 
accuracy of future measurements cannot be underestimated.

\begin{figure}
\centering
\includegraphics[width=8cm,clip]{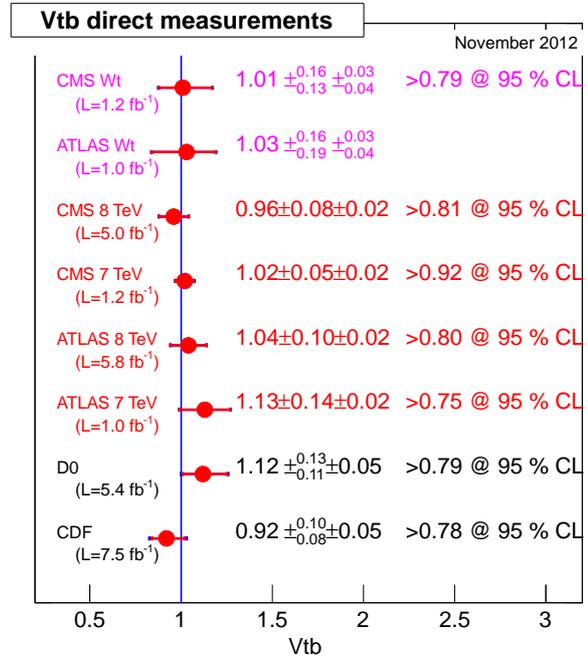}
\caption{All $|V_{tb}|$ measurements using single top cross sections.}
\label{vtb}       % Give a unique label
\end{figure}

Using the current single top cross section measurements, one can compile 
(see 
figure~\ref{vtb}) a list of the direct determination of $|V_{tb}|$. The 
situation is far from being satisfying. 
While the CMS 7 TeV 
measurement~\cite{slhcms} has an uncertainty of $\approx $ 5 \%, all the 
other results (including the most recent one by CMS at 8 TeV) 
have uncertainties at the level of 10\%, 
too large to challenge the SM. 
As the 
LHC has already large statistics, 
these measurements are (mostly) systematics limited. It is a challenge to 
tackle in order to use this process as a tool to search for new physics in 
the top sector.

\section{Acknowledgments}
I would like to thanks the Organizers for inviting me and my colleagues of 
the different experiments for the useful information. Finally I would like 
to remember late Kuni Kondo, my colleague for many years who first 
introduced me to Japan at the PPbar workshop held in KEK in 1993.
%
% BibTeX or Biber users please use (the style is already called in the class, ensure that the "woc.bst" style is in your local directory)
% \bibliography{name or your bibliography database}

\end{document}